\documentclass[10pt, twoside, a4paper]{article}
\begin{document}
\newcommand{\bea}{\begin{equation}}
\newcommand{\eea}{\end{equation}}
\author{N. Bazunova
\footnote{\it University of Tartu, Institute of Pure Mathematics,
Vanemuise 46, 51014 Tartu, Estonia, nadegda@ut.ee}\ ,
A. Borowiec
\footnote{\it University of Wroc{\l}aw,Institute of Theoretical Physics,
plac Maksa Borna 9, PL 50-204 Wroc{\l}aw, Poland,
borow@ift.uni.wroc.pl}\ ,
R.Kerner
\footnote{\it L.P.T.L. - Tour 22, 4-\`eme \'etage, Boite 142,
Universit\'e Paris-VI, 4, Place Jussieu, 75005 Paris,
rk@ccr.jussieu.fr}}
\date{\small To appear in Czechoslovak Journal of  Physics v. 51 (2001)}
\title{Quantum de Rham complex with $d^3 = 0$ differential}
\maketitle
\begin{abstract}
In this work, we construct the de Rham complex with differential
operator $d$ satisfying the $Q$-Leibniz rule, where $Q$ is a
complex number, and the condition $d^3=0$ on an associative
unital algebra with quadratic relations. Therefore we introduce
the second order differentials $d^2x^i$.
In our formalism, besides the usual two-dimensional quantum plane, we
observe that the second order differentials $d^2 x$ and $d^2 y$ generate
either  bosonic or fermionic quantum planes, depending on the choice of the
differentiation parameter $Q$.
\end{abstract}

\section{Introduction}
\indent
Since the discovery of quantum plane by Yu. V. Manin and its possible
applications for the description of deformed or more intricate than usual
symmetries in mathematical physics by Wess and Zumino, an immense activity
followed, especially during the past decade. Quite naturally, after the
purely algebraic properties of those newly discovered spaces have been
quite deeply investigated, and the related quantum groups and Hopf algebras
analyzed and described, the study of analytic properties had followed. This
is why the $q$-deformed algebras have become the next object of many
excellent studies \cite{Carow}, \cite{Madore}, \cite{Kerner}.
\newline
\indent
Parallelly, novel ternary and $Z_3$-graded algebraic structures have been
introduced and investigated \cite{Kerner1}, then generalized to the
$Z_N$-graded case \cite{MDVRK}, \cite{Dubois-Violette}, \cite{MDVITT}.
Thus an important class of $Z_N$-graded differential algebraic structures
has been investigated in an exhaustive manner.

It becomes natural now to combine these two novel and important structures
in order to see whether they can lead to further generalizations of many
useful algebraic and analytic tools such as homology, de Rham complexes,
Hecke and braided algebras, and the like. The aim of this article is to show
the de Rham complex can be generalized for the case of the non-standard
differential satisfying $d^3 = 0$ but with $d^2 \neq 0$.

The paper is organized as follows.
In the first section, we elaborate the general formalism for
differential calculus with the differential operator $d$
satisfying the $Q$-Leibniz rule and the condition $d^3=0$ on an
associative unital algebra with quadratic relations. Supposing
$d^2\ne 0$, we introduce the second order differentials
$d^2x^i$ and find the relations connecting the generators
$x^i$ and $d^2x^j$, $dx^i$ and $d^2x^j$.

In the second section, we find the values of parameter $Q$ from
the commutation relations on the second order differentials $d^2x$
and $d^2y$. For these  values of $Q$ we find that $d^2x$ and $d^2 y$
generate either bosonic or fermionic quantum plane, depending on the
value chosen for $Q$.

In what follows, we shall use the notation
$[k]_Q = 1 + Q + Q^2 +\dots+ Q^{k-1} $;\ we shall denote by $E$ the
identity operator (or matrix) acting in linear space defined by the
context.

\section{General case}

Let $\mathcal{A}$ be an associative unital algebra generated by
variables $x^1,\ x^2, \dots ,x^n$, which satisfy commutation
relations \bea \label{q-pl} x^{i}x^{j}=B_{kl}^{ij}x^{k}x^{l}, \eea where $B$
is a matrix with complex number entries.

Our aim is to generalize de Rham complex by assuming that $d^3=0$
instead of the usual condition $d^2=0$. De Rham complex consists of a
first order differential calculus and it's higher order prolongation.
In our
construction, the first order differential calculus coincides with
Wess-Zumino type differential calculus
on the quantum plane (\ref{q-pl}) \cite{W-Z}.
It is determined by the commutation
relations between the generators $x^i$ and the differential one-forms
$dx^j,\ i,j=1,\ 2, \dots ,n$,
\bea \label{fdc}
x^{i}dx^{j}=C_{kl}^{ij}dx^{k}x^{l},
\eea
satisfying consistency
conditions involving matrices $B$ and $C$:
\begin{eqnarray}
 \label{lincond} (E_{12}-B_{12})(E_{12}+C_{12})\ =\ 0,\\
\label{bcc}
B_{12}C_{23}C_{12}\ =\ C_{23}C_{12}B_{23}.\end{eqnarray}
Where $E_{12}$ denotes the identity operators tensor product,
$E_{12} = I_1 \otimes I_2$, operating in the tensor product of the space
of $1$-forms $d x^i$ with the quantum space generated by $x^k$.

The equation (2) can be interpreted as a definition of a left
${\mathcal{A}}$-module, therefore also a bimodule structure on a free right
$\mathcal{A}$-module generated by the
differentials $dx^j$ (see e.g. \cite{Bor-Kharch} for more
details).

Since we assume $d^2\ne 0$, we must introduce the second order
differentials and
replace the classical Leibniz and eventually generalize the
classical Leibniz rule. The
differential $d$, by means of
which we extend the first order differential calculus, is now supposed
to satisfy the $Q$-Leibniz rule:\
\mbox{$ d(\omega \theta )=d\omega\
\theta +Q^{{\rm{deg}}(\omega )}\omega\ d\theta,
$}\
where
$\omega$ and $\theta$ are differential forms, ${\rm{deg}}(\omega)$ is the
grade of the element $\omega$, and $Q$ is a complex number.
In particular, for ${\rm{deg}}(\omega)={\rm{deg}}(\theta)=0$,
i.e. for a first order calculus, one recovers the standard (undeformed)
Leibniz rule.

The first differentiation of (\ref{fdc}) gives rise to the relations
between the generators $x^i$, the first and second order differentials
$dx^j, \ d^2x^k$:
\bea \label{xd2x}
x^{i}d^{2}x^{j}=C_{kl}^{ij}d^{2}x^{k}x^{l}+(QC_{kl}^{ij}-\delta
_{k}^{i}\delta _{l}^{j})dx^{k}dx^{l}. \eea

Next differentiation gives the commutation relations  between
differentials\ $ dx^{k}$ and $d^{2}x^{l}$ only:
 \bea \label{dxd2x}
([2]_Q\, \delta _{k}^{i}\delta
_{l}^{j}-Q^{2}C_{kl}^{ij})dx^{k}d^{2}x^{l}=([2]_Q\, QC_{kl}^{ij}-\delta
_{k}^{i}\delta _{l}^{j})d^{2}x^{k}dx^{l}. \eea

Finally, we obtain extra relations between differentials
$d^{2}x^{i}$:
\bea \label{d2xd2x}
[3]_Q\, d^{2}x^{i}d^{2}x^{j}=[3]_Q\, Q^{2}C_{kl}^{ij}d^{2}x^{k}
d^{2}x^{l}.
\eea
When $Q$ is not a primitive cubic root of unity, i.e.
$Q \ne e^{\frac{2\pi i}{3}}$, we arrive at the following relations
\bea \label{d2xd2x}
d^{2}x^{i}d^{2}x^{j}=Q^{2}C_{kl}^{ij}d^{2}x^{k}d^{2}x^{l}.
\eea
Therefore we refer to the case $Q=e^{\frac{2\pi i}{3}}$ as
specific because in this case there is no need to introduce new relations
between the generators $d^2 x^i$ (for more general cases,
see \cite{D-VandK}, \cite{D-V}).

The relations (\ref{q-pl} - \ref{d2xd2x}) define a universal
quantum ternary de Rham complex: any other de Rham complex on the
quantum plane (\ref{q-pl}) admitting the first order calculus (\ref{fdc})
can be obtained from this one via a standard quotient construction.
In particular,
we can assume that there exist commutation relations between generators
$x^i$ and second order differentials $d^2x^i$
\bea \label{F-rel}
x^{i}d^{2}x^{j}=F_{kl}^{ij}d^{2}x^{k}x^{l}.
\eea
As a consequence, the second order differentials have to satisfy
the following relations
\bea
d^{2}x^{i}d^{2}x^{j}= Q^{4}F_{kl}^{ij}d^{2}x^{k}
d^{2}x^{l}.
\eea
Again, this requirement allows us to introduce a left
$\mathcal{A}$-module, therefore also a bimodule, structure on a
right free module generated by the second order differentials
$d^2x^j$. Because of this, $F$ should satisfy quadratic consistency
conditions analogous to the condition (\ref{bcc}):
\bea
\label{bff}
B_{12}F_{23}F_{12}=F_{23}F_{12}B_{23}.
\eea
Substituting now (\ref{F-rel}) into (\ref{xd2x}), one finds
\bea
(F^{ij}_{kl}-C_{kl}^{ij})d^{2}x^{k}x^{l}=(QC_{kl}^{ij}-\delta
_{k}^{i}\delta _{l}^{j})dx^{k}dx^{l},
\eea
and using (\ref{dxd2x}), the consistency condition takes on the form :
\bea
E-(Q^2+Q)C+((Q^2+Q)E-Q^3C)Q F=0.
\eea
The last equation reduces, in the generic case $Q = e^{\frac{2 \pi i}{3}}$
to a linear (cf. (\ref{lincond})) Wess-Zumino-like condition on the
matrices $C$ and $F$:
\bea \label{cqf}
(E+C)(E-Q F)=0.
\eea

Following the well known Wess-Zumino method, we can now resolve the
consistency conditions (\ref{bff}) and (\ref{cqf}). To this end let us assume that
a Hecke $R$-matrix is given, i.e. the matrix $R$ satisfying the braid
relation :
\bea
R_{12}R_{23}R_{12}=R_{23}R_{12}R_{23}
\eea
together with the second-order minimal polynomial condition
\bea
(R-\mu E)(R+\lambda E)=0.
\eea
Rewriting it in the form
\bea
(E-\frac{1}{\mu} R)(E+\frac{1}{\lambda} R)=0,
\eea
one immediately sees that $B=\frac{1}{\mu} R$,\ $C=\frac{1}{\lambda} R$
and\ $F=\frac{Q^2}{\mu}R$
are the solution of the consistency conditions
(\ref{lincond}), (\ref{bcc}) and (\ref{bff}), (\ref{cqf}).
These can be further generalized for non-Hecke $R$-matrices
(cf. \cite{Hlavaty}).

More generally, having matrices $B$ and $C$ as a solution of the
consistency conditions (\ref{lincond}, \ref{bcc}), one can set
$F=Q^2B$ provided that $B$ commutes with $C$ and the braid
relation (\ref{bff}) is satisfied.

\section{Two-dimensional quantum plane}

In this section, we shall construct the de Rham complex
with $d^3=0$ on the two-dimensional quantum plane.

As it is well known, the quantum plane $xy = q\,y x$, where $q$ being
a complex deformation
parameter, is determined by an $R$-matrix
\begin{displaymath}
B=\frac{1}{q}\widehat{R},\ \ \textrm{where} \ \ \widehat{R}=
\left(
\begin{array}{cccc}
q & 0 & 0 & 0 \\
0 & q-q^{-1} & 1 & 0 \\
0 & 1 & 0 & 0 \\
0 & 0 & 0 & q
\end{array}
\right)
\end{displaymath}
satisfying the braid relation.
There are two infinite and non-equivalent families of covariant
first order differential calculus on this plane (parameterized by
a complex parameter $r$), which are characterized by means of
matrices $C_1$ and $C_2$ which define the commutation relations
(\ref{fdc})(cf. \cite{Pusz-Wor})
\begin{displaymath}
C_1=
\left(
\begin{array}{cccc}
r & 0 & 0 & 0 \\
0 & r-1 & q & 0 \\
0 & \frac{r}{q} & 0 & 0 \\
0 & 0 & 0 & r
\end{array}
\right)
\ \ \textrm{and} \ \
C_2=
\left(
\begin{array}{cccc}
r & 0 & 0 & 0 \\
0 & 0 & q r & 0 \\
0 & \frac{1}{q} & r-1 & 0 \\
0 & 0 & 0 & r
\end{array}
\right).
\end{displaymath}
As a matter of fact, the two matrices $C_1$ and $C_2$ are not
really indepandent: one can be obtained from another if we
substitute $x$ by $y$ and simultaneously $q$ by $q^{-1}$ and
vise versa.
They define a generalization of the first-order differential calculi
obtained by Wess and Zumino \cite{W-Z}. In fact, we get a Wess-Zumino
first order
differential calculi if $r = q^2$ for the matrix $C_1$ and $r = 1/q^2$
for the matrix $C_2$ (cf. \cite{Soni}).

Following the general formalism elaborated above, from
(\ref{d2xd2x}) we get two sorts
of commutation relations for generators $d^2x, \ d^2y$.
Considering the obtained relations as equations
with respect to the parameter of differentiation $Q$, we get the
set of values $Q$: $\{ e^{\frac{2\pi i}{3}},\ e^{\frac{4\pi
i}{3}},\ \pm \frac{1}{\sqrt{r}},\ i\}$.

If $Q=e^{\frac{2\pi i}{3}}$ or $Q=e^{\frac{4\pi i}{3}}$,
then $d^2x$ and $d^2y$ can not satisfy any particular binary
relations.

If $Q=\pm \frac{1}{\sqrt{r}}$, then we have the commutation
relations without the parameter $r$:
\bea
\begin{array}{ccc}
d^{2}x\,d^{2}x=d^{2}x\,d^{2}x,&&
d^{2}x\,d^{2}y=q\,d^{2}y\,d^{2}x,\nonumber\\
d^{2}y\,d^{2}x=q^{-1}\,d^{2}x\,d^{2}y,&&
d^{2}y\,d^{2}y=d^{2}y\,d^{2}y\nonumber
\end{array}
\eea
for both choices of matrices $C_1$ and $C_2$. It means that the
generators $d^2x$ and $d^2y$ define the quantum plane
$d^{2}x\,d^{2}y = q\,d^{2}y\,d^{2}x$, which is like
the given quantum plane $xy = q\,y x$. Both quantum planes are
preserved by the
action of quantum group $GL_q(2)$, determined by generators
$\alpha,\ \beta,\gamma,\ \delta$ satisfying the commutation
relations:
\bea
\begin{array}{ccccc}
\alpha\beta=q\beta\alpha,&&\alpha\gamma=q\gamma\alpha,&&
\beta\gamma=\gamma\beta,\nonumber\\
\gamma\delta=q\delta\beta,&&\beta\delta=q\delta\beta,&&
\alpha\delta-\delta\alpha=(q-q^{-1})\beta\gamma.\nonumber\\
\end{array}
\eea

The parameter $r$ does not vanish when $Q = i = e^{\frac{i \pi}{2}}$, but
in this case the matrices $C_1$ and $C_2$ define two distinct quantum planes.
These quantum planes are preserved by the action of two different quantum
groups $GL_{r,q} (2)$.

In fact, if the first order differential calculus is determined
by the matrix $C_1$, then we have the following commutation
relations between $d^2x,\ d^2y$:
\bea
\begin{array}{ccc}
d^{2}x\,d^{2}x=-r\,d^{2}x\,d^{2}x,&&
d^{2}x\,d^{2}y=-\frac{q}{r}\,d^{2}y\,d^{2}x,\nonumber\\
d^{2}y\,d^{2}x=-\frac{r}{q}\,d^{2}x\,d^{2}y,&&
d^{2}y\,d^{2}y=-r\,d^{2}y\,d^{2}y.\nonumber
\end{array}
\eea
From first and fourth relations, it follows that
$(d^2x)^2=(d^2y)^2=0$. The quantum plane determined by
relations
$d^{2}x\,d^{2}y=-\frac{q}{r}\,d^{2}y\,d^{2}x,\ (d^2x)^2=(d^2y)^2=0$
is preserved by action of quantum group $GL_{r,q}$ whose generators
$\alpha,\ \beta,\gamma,\ \delta$ satisfy the
commutation relations:
\bea
\begin{array}{ccccc}
\alpha\beta=\frac{r}{q}\,\beta\alpha,&&\alpha\gamma=q\gamma\alpha,&&
\frac{r}{q}\,\beta\gamma=q\gamma\beta,\nonumber\\
\beta\delta=\frac{r}{q}\,\delta\beta,&&\beta\delta=q\delta\beta,&&
\alpha\delta-\delta\alpha=(q-(\frac{r}{q})^{-1})\,\gamma\beta=
(\frac{r}{q}-q^{-1})\beta\gamma.
\end{array}
\eea

In the case of the first order differential calculus determined
by matrix $C_2$, we get the  commutation relations:
\bea
\begin{array}{ccc}
d^{2}x\,d^{2}x=-r\,d^{2}x\,d^{2}x,&&
d^{2}x\,d^{2}y=-qr\,d^{2}y\,d^{2}x,\nonumber\\
d^{2}y\,d^{2}x=-(qr)^{-1}\,d^{2}x\,d^{2}y,&&
d^{2}y\,d^{2}y=-r\,d^{2}y\,d^{2}y, \nonumber
\end{array}
\eea
which define the quantum plane
$d^{2}x\,d^{2}y=-qr\,d^{2}y\,d^{2}x,\ (d^2x)^2=(d^2y)^2=0$.
This quantum plane is preserved by action of quantum group
$GL_{r,q}(2)$, generated by $\alpha,\ \beta,\gamma,\ \delta$
satisfying a different set of commutation relations:
\bea
\begin{array}{ccccc}
\alpha\beta=\frac{1}{rq}\,\beta\alpha,&&\alpha\gamma=q\gamma\alpha,&&
\frac{1}{q}\,\beta\gamma=rq\gamma\beta,\nonumber\\
\gamma\delta=\frac{1}{rq}\,\delta\beta,&&\beta\delta=q\delta\beta,&&
\alpha\delta-\delta\alpha=(q-rq)\,\gamma\beta=
(\frac{1}{q}-\frac{1}{rq})\beta\gamma.
\end{array}
\eea

\bigskip
{\small N.B. and A.B. wish to thank for hospitality the
Laboratoire LPTL where this paper has been written.
N.B. is very grateful to V. Abramov and to the organizers of
the present Colloquium for financial support, and acknowledges
the financial support of Estonian Science Foundation under the grants
No.1134 and No.4515.
\bigskip


\begin{thebibliography}{9}
\bibitem{Carow} U. Carow-Watamura, S. Watamura: Int. J. Mod. Phys. A {\bf13},
No.19, 1998, 3235-3243 .
\bibitem{Madore}   B.L. Cherchiai, R. Henterding, J. Madore, J. Wess:
Eur. Phys. J. {\bf 8}, No.3, 1998, p.547-558.
\bibitem{Kerner} R. Kerner:{\it $Z\sb 3$-graded exterior differential
calculus and gauge theories of higher order}
Lett. Math. Phys. {\bf 36}, No.4, 1996, p.441-454; math-ph/0004032.
\bibitem{MDVRK}
M. Dubois-Violette, R. Kerner:  Acta Math. Univ. Comen., New Ser.
{\bf65},No.2, 1996, p.175-188; q-alg/9608026.
\bibitem{Kerner1} R.Kerner: in {\it the proceedings of the Conference ICGTMP
"Group-23"}, Dubna, Russia, July 30 - August 6, 2000;
math-ph/0011023.
\bibitem{MDVITT} M.Dubois-Violette and I.T.Todorov: Lett. Math. Phys. {\bf48},
No.4, 1999, p.323-338; hep-th/9704069.
\bibitem{Dubois-Violette} M.Dubois-Violette: Czech. J. Phys. {\bf46},
No.12, 1996, p.1227-1233; q-alg/9609012.
\bibitem{W-Z} J. Wess, B. Zumino: Nucl. Phys. B, Proc. Suppl.
{\bf18B} 1990, p.302-312.
\bibitem{Bor-Kharch} V. K. Kharchenko, A. Borowiec,
{\it Questions of algebra and logic}, Izdatel'stvo Instituta Matematiki SO
RAN. Tr. Inst. Mat. Im. S. L. Soboleva SO RAN. 30, Novosibirsk,
1996, p.164-185 (in Russian).
\bibitem{Hlavaty} L. Hlavaty: J. Phys. A, Math. Gen. {\bf 25}
1992, p.485-494.
\bibitem{Pusz-Wor} W. Pusz, S. Woronowicz: Rep. Math. Phys. {\bf 27}, No.2,
1989, p.231-257.
\bibitem{Soni} S.K. Soni: J. Phys. A, Math. Gen. {\bf24}, 1991,
p.169-174.
\bibitem{D-VandK} Dubois-Violette, M., Kerner, R.:
Acta Math. Univ. Comen., New Ser. {\bf 65}, No.2, 1996, p.175-188.
\bibitem{D-V}  Dubois-Violette, M.: K-Theory {\bf 14}, No.4, 1998,
p.371-404.
\end{thebibliography}
\end {document}